\documentclass[a4,aps,prc,twocolumn,12pt,superscriptaddress]{revtex4}
\usepackage{graphicx}
\usepackage{times}
\usepackage[latin1]{inputenc}
\usepackage{amssymb}

\begin{document}

\title{\large DYNAMICS AND THERMODYNAMICS OF PHASE TRANSITION IN HOT NUCLEI}

\author{M. F. Rivet} 
\email[\footnotesize e-mail:]{rivet@ipno.in2p3.fr}
\affiliation{\rm Institut de Physique Nucl\'eaire, IN2P3-CNRS, F-91406 
Orsay Cedex,  France.} 
%\thanks{}
\author{N.~Bellaize}
\affiliation{\rm LPC, IN2P3-CNRS, ISMRA et Université, F-14050 Caen Cedex, France.}
\author{B.~Borderie}
\affiliation{\rm Institut de Physique Nucl\'eaire, IN2P3-CNRS, F-91406 
Orsay Cedex,  France.}
\author{R.~Bougault}
\affiliation{\rm LPC, IN2P3-CNRS, ISMRA et Université, F-14050 Caen Cedex, France.}
\author{A. Chbihi}
\author{J.D. Frankland}
\author{B.~Guiot}
\affiliation{\rm GANIL, CEA et IN2P3-CNRS, B.P.~5027, F-14076 Caen Cedex, France.}
\author{O.~Lopez}
\author{N.~Le~Neindre}
\affiliation{\rm LPC, IN2P3-CNRS, ISMRA et Université, F-14050 Caen Cedex, France.}
\author{M.~P\^arlog}
\author{G.~T\u{a}b\u{a}caru} 
\affiliation{\rm Nat. Inst. for Physics and Nuclear Engineering, 
Bucharest-M\u{a}gurele, Romania.\\} 
\author{J.P.~Wieleczko}
\affiliation{\rm GANIL, CEA et IN2P3-CNRS, B.P.~5027, F-14076 Caen Cedex, France.}
\author{G.~Auger}
\affiliation{\rm GANIL, CEA et IN2P3-CNRS, B.P.~5027, F-14076 Caen Cedex, France.}
\author{Ch.O.~Bacri}
\affiliation{\rm Institut de Physique Nucl\'eaire, IN2P3-CNRS, F-91406 Orsay Cedex,
 France.}
\author{B.~Bouriquet}
\affiliation{\rm GANIL, CEA et IN2P3-CNRS, B.P.~5027, F-14076 Caen Cedex, France.}
\author{A.M.~Buta}
\affiliation{\rm LPC, IN2P3-CNRS, ISMRA et Université, F-14050 Caen Cedex, France.}
\author{J.~Colin}
\author{D.~Cussol}
\affiliation{\rm LPC, IN2P3-CNRS, ISMRA et Université, F-14050 Caen Cedex, France.}
\author{R.~Dayras}
\affiliation{\rm DAPNIA/SPhN, CEA/Saclay, F-91191 Gif sur Yvette, France.}
\author{N.~de~Cesare}
\affiliation{\rm Dip. di Scienze Fisiche e Sez. INFN, Univ. di Napoli
``Federico II'', Napoli, Italy.}
\author{A.~Demeyer}
\affiliation{\rm Institut de Physique Nucléaire, IN2P3-CNRS et Universit\'e
F-69622 Villeurbanne, France.}
\author{D.~Dor\'e}
\affiliation{\rm DAPNIA/SPhN, CEA/Saclay, F-91191 Gif sur Yvette, France.}
\author{D.~Durand}
\affiliation{\rm LPC, IN2P3-CNRS, ISMRA et Université, F-14050 Caen Cedex, France.}
\author{E.~Galichet}
\affiliation{\rm Institut de Physique Nucl\'eaire, IN2P3-CNRS, F-91406 Orsay Cedex,
 France.}
\affiliation{\rm Conservatoire National des Arts et Métiers, F-75141 Paris
Cedex 03.}
\author{E.~Gerlic}
\author{D.~Guinet}
\affiliation{\rm Institut de Physique Nucléaire, IN2P3-CNRS et Universit\'e
F-69622 Villeurbanne, France.}
\author{S.~Hudan}
\affiliation{\rm GANIL, CEA et IN2P3-CNRS, B.P.~5027, F-14076 Caen Cedex, France.}
\author{G.~Lanzalone}
\altaffiliation[Permanent address: ]{Laboratorio Nazionale del Sud, 
Via S. Sofia 44, I-95123 Catania, Italy.}
\affiliation{\rm Institut de Physique Nucl\'eaire, IN2P3-CNRS, F-91406 Orsay Cedex,
 France.}
\author{P.~Lautesse}
\affiliation{\rm Institut de Physique Nucléaire, IN2P3-CNRS et Universit\'e
F-69622 Villeurbanne, France.}
\author{F.~Lavaud}
\affiliation{\rm Institut de Physique Nucl\'eaire, IN2P3-CNRS, F-91406 Orsay Cedex,
 France.}
\author{J.L.~Laville}
\affiliation{\rm GANIL, CEA et IN2P3-CNRS, B.P.~5027, F-14076 Caen Cedex, France.}
\author{J.F.~Lecolley}
\affiliation{\rm LPC, IN2P3-CNRS, ISMRA et Université, F-14050 Caen Cedex, France.}
\author{R.~Legrain}
\thanks{deceased}
\author{L.~Nalpas}
\affiliation{\rm DAPNIA/SPhN, CEA/Saclay, F-91191 Gif sur Yvette, France.}
\author{J.~Normand}
\affiliation{\rm LPC, IN2P3-CNRS, ISMRA et Université, F-14050 Caen Cedex, France.}
\author{P.~Paw{\l}owski}
\author{E.~Plagnol}
\affiliation{\rm Institut de Physique Nucl\'eaire, IN2P3-CNRS, F-91406 Orsay Cedex,
 France.}
\author{E.~Rosato}
\affiliation{\rm Dip. di Scienze Fisiche e Sez. INFN, Univ. di Napoli
``Federico II'', Napoli, Italy.}
\author{R.~Roy}
\affiliation{\rm Laboratoire de Physique Nucléaire, Université Laval,
Québec, Canada.}
\author{J.C.~Steckmeyer}
\author{B.~Tamain}
\author{E.~Vient}
\affiliation{\rm LPC, IN2P3-CNRS, ISMRA et Université, F-14050 Caen Cedex, France.}
\author{M.~Vigilante}
\affiliation{\rm Dip. di Scienze Fisiche e Sez. INFN, Univ. di Napoli
``Federico II'', Napoli, Italy.}
\author{C.~Volant}
\affiliation{\rm DAPNIA/SPhN, CEA/Saclay, F-91191 Gif sur Yvette, France.}

\collaboration{INDRA collaboration} \noaffiliation

%\date{\today}

\begin{abstract}
\begin{center} \textit{\normalsize Abstract} \end{center}
\begin{minipage}{14.7cm}
{\footnotesize The dynamics and thermodynamics of phase transition in hot 
nuclei are
studied through experimental results on multifragmentation of heavy systems
(A$\geq$ 200) formed in central heavy ion collisions. Different signals
indicative of a phase transition studied in the INDRA collaboration 
are presented and their consistency is stressed.} \end{minipage}
\end{abstract}

\maketitle

\section{ INTRODUCTION }

\vspace*{-5mm} Liquid-gas type phase transitions are commonly observed in 
systems with short-range repulsive and longer-range attractive forces,
such as macroscopic fluids with the van der Waals interaction. The nuclear 
equation of state is very similar to that of non-ideal gases, 
which allows to foresee the existence of different phases of nuclear matter.
But can one define phase transitions in  ``small'' objects, when the
dimension is of the same order of magnitude as the
range of the force which binds them? In this sense nuclei as well as 
clusters of galaxies are small. Recent works state that statistical mechanics 
based on the Boltzmann' definition of entropy ($S=k.\log{W}$) allows to 
define phase transitions in small systems~\cite{Gro01}.
 The nucleus at zero or moderate temperature, because of its quantal nature, 
 is assimilated to a liquid phase. A nuclear gas phase was characterized, 
 for instance, through the properties of vaporised quasi-projectiles 
 from 95~AMeV Ar+Ni reactions, fully predicted by modelling a 
 van der Waals gas of fermions and bosons in thermodynamic 
 equilibrium~\cite{Bor99}. The liquid-gas coexistence region was thus 
 naturally connected to multifragmentation, break-up of a nuclear 
 system in several fragments of various sizes.

Most of the works on nucleus phase transition relied on static properties 
such as the charge (or mass) partitions measured in nuclear reactions.
The observation of a power law on a limited range of fragment charge 
measured in inclusive experiments was claimed to be consistent with a 
critical behaviour, according to  Fisher's droplet model~\cite{Fis67}, 
and also to percolation models~\cite{Ell00}.
 Recent and well selected data from the ISIS collaboration were analysed 
 in this framework, and a liquid-gas coexistence curve was 
 derived~\cite{Ell01,Mor01}. 
In the same line, critical exponents were reported to have measured values in
agreement with those of a liquid-gas model~\cite{Ell98,MDA99}. 

Caloric curves relate the energy and the temperature of nuclei. The 
observation of caloric curves showing a bending or a plateau,
similar to what is observed for the boiling of water, was claimed as 
a proof of the nuclear phase transition~\cite{Poc95}.

A new signal of phase transition of nuclei was found in 
the observation of a negative branch of the microcanonical
heat capacity for Au quasi-projectiles excited between 3 and 6
AMeV~\cite{MDA00}.

This paper presents a compilation of the results which deal with 
\emph{central collisions} between heavy ions (A$_{proj}$ + A$_{tgt} >$ 200),
obtained so far in the INDRA collaboration. Most of them will be 
detailed in other contributions to this conference. Different signals
which may be connected with the occurrence of a phase transition in nuclei
were found, each of them bringing a small stone to the construction of a 
consistent framework.
%\subsection{}
%\subsubsection{}

\section{PHASE TRANSITION: DYNAMICS AND THERMODYNAMICS}

\vspace*{-3mm}
\subsection{\textmd{\textit{Thermodynamics aspect}}}\label{thermo}

\begin{figure}[!hbt]
\includegraphics*[scale=0.8]{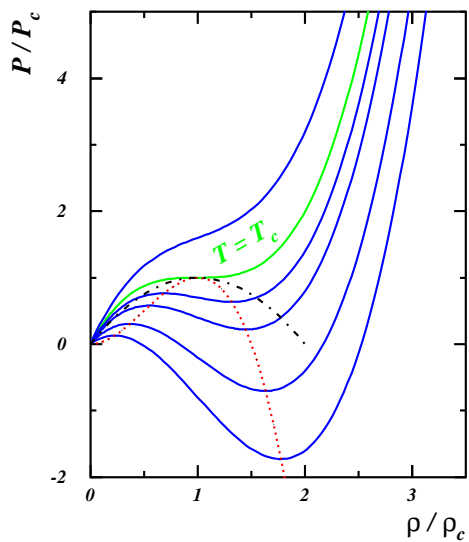}
\vspace*{-10mm} \caption{\footnotesize 
Equation of state relating the pressure and the
density (normalised to the critical values) in  nuclear matter.
The curves represent isotherms. The dashed-dotted line is the
coexistence line and the dotted line the spinodal line. }\label{pv} 
\end{figure}

\begin{figure}[!hbt]
\includegraphics*[scale=0.9]{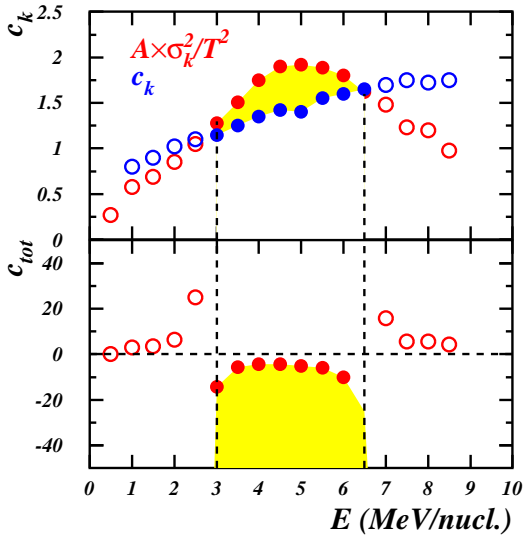}
\vspace*{-10mm} \caption{\footnotesize 
Variation versus energy of the kinetic fluctuations and heat capacity 
(top) and of the total heat capacity (bottom).}
\label{fluccneg} \end{figure}

\vspace*{-3mm} As exemplified in fig.~\ref{pv}, where isotherms are drawn 
in a pressure-density plane, the EOS of nuclear matter resembles 
that of a van der Waals gas. 
A region of negative incompressibility, bordered by  the spinodal
line, and located inside the coexistence region, appears in this diagram.
This spinodal region is a zone of mechanical instability, where an
increase of pressure leads to an expansion of the system;
on entering this zone the system will tend to
recover thermodynamical equilibrium by separating into two coexisting phases,
a low-density (gas) phase and a higher density (liquid) phase. Note that the
time for the system to go from homogeneity to inhomogeneity is finite, in
other words a phase transition has a dynamics (see next subsection).
The spinodal region is also thermally unstable, as the heat capacity,
$c=\mathrm{d}E/\mathrm{d}T$, is negative inside this zone. According to
this formula, it seems easy to evidence a negative heat capacity,
which should appear, in a microcanonical frame, as a backbending
of the caloric curve. It is not so simple in the case of
nuclei, as their temperature, energy, density
can only be modified by a more or less violent nuclear collision;
one is dealing with isolated systems without external constraints,
which eventually induce a phase transition in the course of their
evolution  in a (temperature, pressure, energy) space. The followed path
is not known and there is no \textit{a priori} reason why the resulting
caloric curve would have a plateau, or a backbending~\cite{Cho00}.
Another and more robust signal was
proposed to recognise a phase transition, abnormally large fluctuations of the
kinetic part, $E_k$ of the energy ($E = E_k + E_{pot}$)~\cite{Cho99}.
In this case the heat capacity can be expressed as
$c_{tot} = c^2_k / (c_k-A\sigma _k^2/T^2)$, with
$c_k = \mathrm{d}\langle E_k \rangle/ \mathrm{d}T$. When the
normalised fluctuations exceed the canonical value $c_k$, the heat capacity 
becomes negative, as depicted in fig.~\ref{fluccneg}.
These variables may be determined in an experiment, provided that one can
reconstruct the partition present at the ``freeze-out'' stage, defined as
the instant when nuclear interaction between the present species vanishes.

Finally  this static statistical physics approach supposes that the
ensemble of collisions considered constitutes a statistical ensemble. This
ensemble is characterized by global variables reflecting the pertinent
information resulting from the dynamics of the collisions or (and) the
sorting of the events. Statistical physics concepts can be safely used
providing that there is adequacy between the ensemble
treated and the applied formalism~\cite{Gul01}.

\vspace*{-5mm}
\subsection{\textmd{\textit{Dynamics of the phase transition}}}\label{dynamic}

\vspace*{-3mm} Is it possible to explore the fugitive dynamics of the 
nuclear phase transition? A positive answer was recently given in 
the INDRA collaboration.
Heavy nuclear systems were produced at the same energy ($\sim$ 7 AMeV),
 but with very different total masses (393 and 248), through central
collisions between 36 AMeV Gd and U and 32 AMeV Xe and Sn; the resulting 
measured charge distributions superimposed 
while the fragment multiplicities scaled as the total system
masses~\cite{Riv98}. While this indicates the dominance of phase space
effects, with stochastic fragment production governed by a Boltzmann factor,
it may also sign the presence of bulk instabilities inducing density
fluctuations, and finally leading to the spinodal decomposition of the
nuclear system. And indeed spinodal decomposition is, together with
nucleation, one of the processes describing the dynamics of the liquid-gas
phase transition in nuclear matter~\cite{Idi94}. 
\begin{figure}[!hbt]
\includegraphics*[scale=0.6]{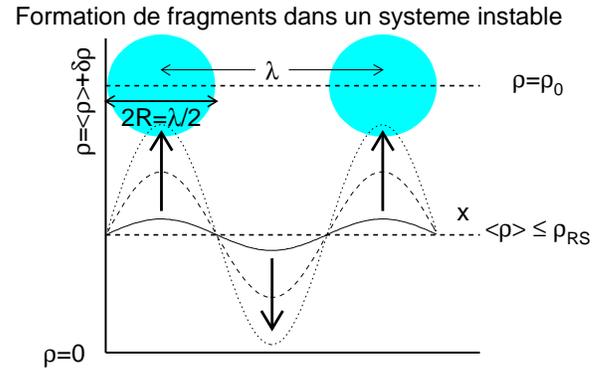}
\vspace*{-10mm} \caption{\footnotesize Density fluctuations and fragment 
formation in the spinodal region~\protect\cite{Fra98}.} \label{flucdens}
\end{figure}
The most unstable mode(s)
in the system induce homogeneous density fluctuations which will be
amplified by the mean field with a characteristic time which decreases with
the temperature of the system. Locally zones close to normal
density may appear, leading to the formation of fragments whose radii and
relative distances depend on the wavelength of the mode, as schematized in
fig.~\ref{flucdens}.
In this scenario, one expects a narrow fragment charge distribution, loosely
dependent on the total charge of the system provided it is large enough. 
In finite size hot nuclear systems several modes are equally probable,
may beat, all effects which blur the initial picture and lead to much 
broader charge distributions. 
\begin{figure}[!hbt]
\includegraphics*[scale=0.9]{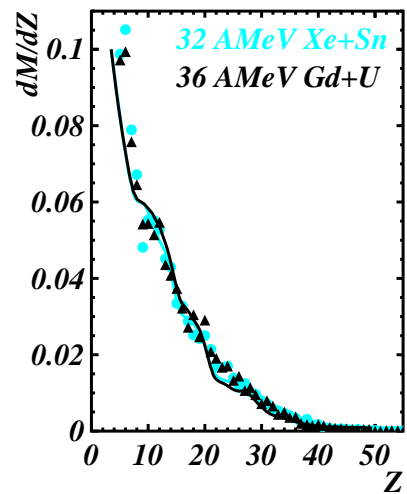}
\vspace*{-10mm} \caption{\footnotesize Charge distributions from central 
heavy ion collisions. The lines show the results of stochastic mean 
field simulations while the symbols display the experimental results. 
Adapted from~\protect\cite{Fra01}.}
\label{dzbob} \end{figure}
This is exemplified in fig.~\ref{dzbob}, where the lines
display the charge distributions obtained through a complete stochastic mean
field simulation of central nuclear collisions between heavy ions, the
Brownian one-body dynamics (Bob): the mean field is complemented by
a Brownian force in the kinetic equations, which mimic a stochastic
collision term. For the two reactions considered here,
36 AMeV Gd+ U, and 32 AMeV Xe+Sn, the system is driven into the spinodal 
region, where it multifragments through spinodal decomposition~\cite{Gua96}.
The de-excitation of the hot fragments so formed is followed
simultaneously with their propagation in their Coulomb field. Finally the
simulated events (Bob events) are passed through a software replica of the
INDRA detector. Obviously the final distributions (lines in 
fig.~\ref{dzbob}) are far from narrow, being
continuously decreasing with increasing Z without any visible peak. 
But they perfectly reproduce the experimental distributions measured in 
the two systems quoted above (symbols in fig.~\ref{dzbob}), 
particularly the independence of the charge 
distribution against the mass of the system is recovered~\cite{Fra01}. It is
worthwhile to mention that kinetic properties of the outgoing channel are
also well accounted for, particularly the fragment-fragment velocity
correlations~\cite{TabBol00}.

\begin{figure*}[!hbt]
\includegraphics*[scale=0.8]{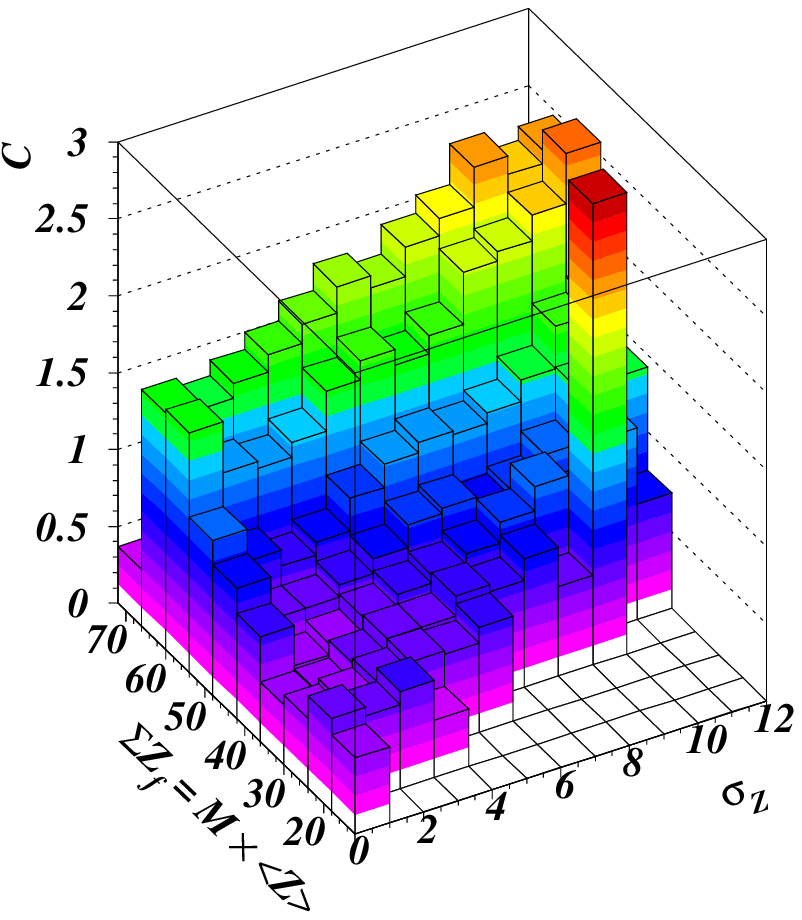}
\includegraphics*[scale=0.8]{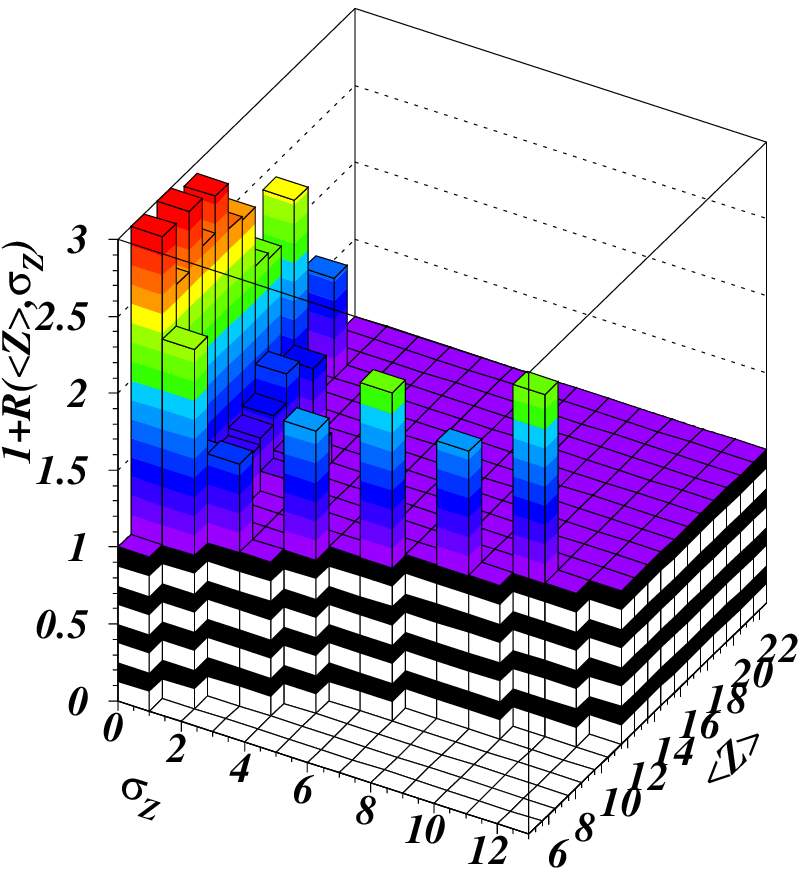}
\vspace*{-8mm} \caption{\footnotesize Charge correlation function for 
events generated by stochastic mean field simulations of central 32 AMeV 
Xe+Sn collisions, without (left) and with (right) charge conservation 
constraint on the uncorrelated yields. Note the different orientation of 
the two figures. The charge axes are $M_f \times Z_{av}$ on the left, 
and $Z_{av}$ on the right, while the $\sigma_Z$ axes are the same. 
In both cases events with 3 to 6 fragments are included.}
\label{corbob}
\end{figure*}
These simulations show that the phase transition, or the spinodal 
decomposition process can not be evidenced by a single variable like a
charge distribution. It might however well be that a few events kept a 
memory of the process, through a preferred break-up in equal-sized fragments. 
The charge correlation function defined as:
$C = Y_{cor}(Z_{av},\sigma _Z) /Y_{uncor}(Z_{av},\sigma_Z)$ is very
efficient to enlighten any extra production of rare event types~\cite{Mor96},
as it makes use of the full per-event information  through two variables,
the average charge $Z_{av}$ and the standard deviation $\sigma _Z$.
Equal-sized fragment partitions (small $\sigma_Z$), were looked for
in Bob events from 32 AMeV Xe+Sn central collisions.
The variable $Z_{av}$ was replaced by $M_f \times Z_{av}$ as events
with different fragment multiplicities $M_f$ were mixed to
improve statistics~\cite{Bor01}.
The uncorrelated yield, $Y_{uncor}(Z_{av},\sigma_Z)$,
obtained  in ref~\cite{Mor96,Bor01} by making pseudo-events from fragments
belonging to different events of the same sample, may also be analytically
calculated from the charge distribution~\cite{Des01}, in order to make
its statistical error negligible everywhere.
The correlation function shown in fig.~\ref{corbob} (left) uses the
analytical $Y_{uncor}$ and the standard deviation of the measure:
$\sigma_Z = (\sum_{i=1}^{M_f}{(Z_i -Z_{av})^2}/M_f)^{1/2}$,
at variance with ref~\cite{Mor96,Bor01}, 
The general trend is a decrease of the correlation function with decreasing
$\sigma_Z$, as expected from the broad charge distribution, except in the
smaller bin, $\sigma_Z<1$, where it increases again. For
$36 < M_f \times Z_{av} < 60$ this increase is statistically significant,
with a confidence level larger than 98\%. But only 0.6\% of the events
contribute to these enhanced partitions, while
all simulated events multifragment through spinodal decomposition.
Due to this small number one may wonder about the real meaning of these 
peaks at small $\sigma_Z$, because their
statistical significance must be weighted against a ``background'', 
corresponding to the extrapolation of the trend of the measured 
correlation function for decreasing $\sigma_Z$. 
This extrapolation was empirically performed with an exponential 
function~\cite{Bor01}. The method was recently improved by
building the uncorrelated yield following an independent emission model 
under the constraint of
charge conservation of the \emph{total system}~\cite{Des01}. The result,
on the same Bob event sample, is shown in the right part of fig.~\ref{corbob}.
All correlations due to charge conservation are now suppressed, and the
correlation function is equal to 1 wherever no additional physical
correlation is present. To make the picture stronger, the peaks with a
significance lower than 95\% have been flattened out in fig.~\ref{corbob}
(right). Again one observes peaks for small values of $\sigma_Z$, the 
percentage of events concerned is 0.7\%, which fully validates the 
published results~\cite{Bor01}.

In short simulations of central heavy ion collisions show that fingerprints 
of the spinodal decomposition of a nuclear system can be found, namely a
weakly but unambiguously enhanced proportion of equal-sized fragment 
partitions, and thus evidence its phase transition. 
A statistical model, like any model
just based on phase space, will not show this type of phenomenon~\cite{Des01}.

\section{EXPERIMENTAL INDICATIONS OF MULTIFRAGMENTATION IN THE SPINODAL REGION
} \label{A250}

\vspace*{-5mm} Two reactions were studied by the INDRA collaboration 
which lead to systems with close total mass and charge 
(Z$\sim$105, A$\sim$250). The symmetric Xe+Sn system was studied 
between 25 and 50 AMeV, and a class of events characterized as 
multifragmentation of ``fused'' systems
was recognized at and above 32 AMeV. Event selection was performed on
completeness and event shape criteria, and the multifragmentation of
deformed objects was not considered in these samples~\cite{Fra01,Bou01}.
The asymmetric Ni+Au system was studied from 32 to 90 AMeV. Owing to
the thresholds of INDRA, and to the smaller c.m. velocity for this
asymmetric entrance channel, the number of complete events is very small.
Methods selecting classes of events were developed, less stringent
on completeness requirements. Two of these methods were used for isolating 
events samples of central collisions with close distributions 
of global observables: the Principal Component Analysis 
method~\cite{BelBorm01,LopBorm01}) for the sample on which the heat 
capacity was determined  and the Discriminant Factorial Analysis 
method~\cite{Des00}) for the sample used for charge correlations.

\vspace*{-5mm}
\subsection{\textmd{\textit{The symmetric entrance channel}}}

\begin{figure}[!hbt]
\includegraphics*[scale=0.6]{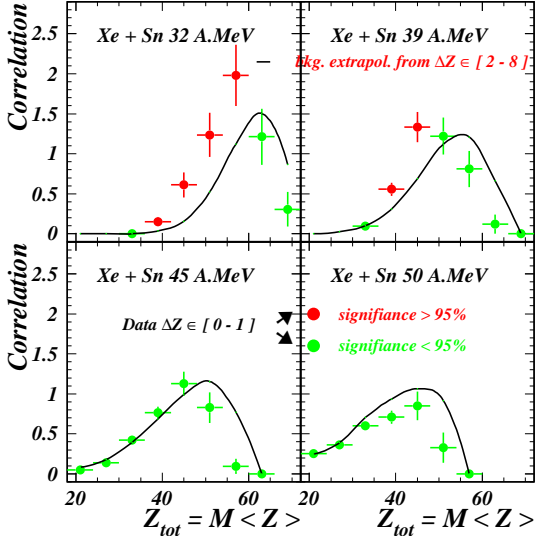}
\vspace*{-4mm} \caption{\footnotesize Charge correlation functions 
(without charge conservation constraint) for central collision
events between Xe and Sn. The lines indicate the
extrapolated background. The darker points correspond to event
partitions enhanced with a high significance. From~\cite{BBorm01}} 
\label{corxesn} \end{figure}

\vspace*{-3mm} 
The heat capacity and the charge correlations were both examined in the same
samples of Xe+Sn central collisions. 
Charge correlations are presented in fig.~\ref{corxesn}, for the
four incident energies~\cite{BBorm01}. The correlation functions shown here
are calculated without the charge conservation constraint, and the figure
only presents them for the first  bin $\sigma_Z \in [0-1]$.
The data corresponding to an enhanced production of equal-sized fragment
partitions with a high confidence level are indicated by darker points. 
This enhancement is only observed at 32 and 39 AMeV, and correspond to 0.2 
and 0.4\% of the selected samples.
These percentages are close to those observed in the Bob simulations, and
may be interpreted in stating that the dynamics of multifragmentation 
occurs by spinodal decomposition in all the selected central collisions.

\begin{figure}[!hbt]
\includegraphics[width=4cm]{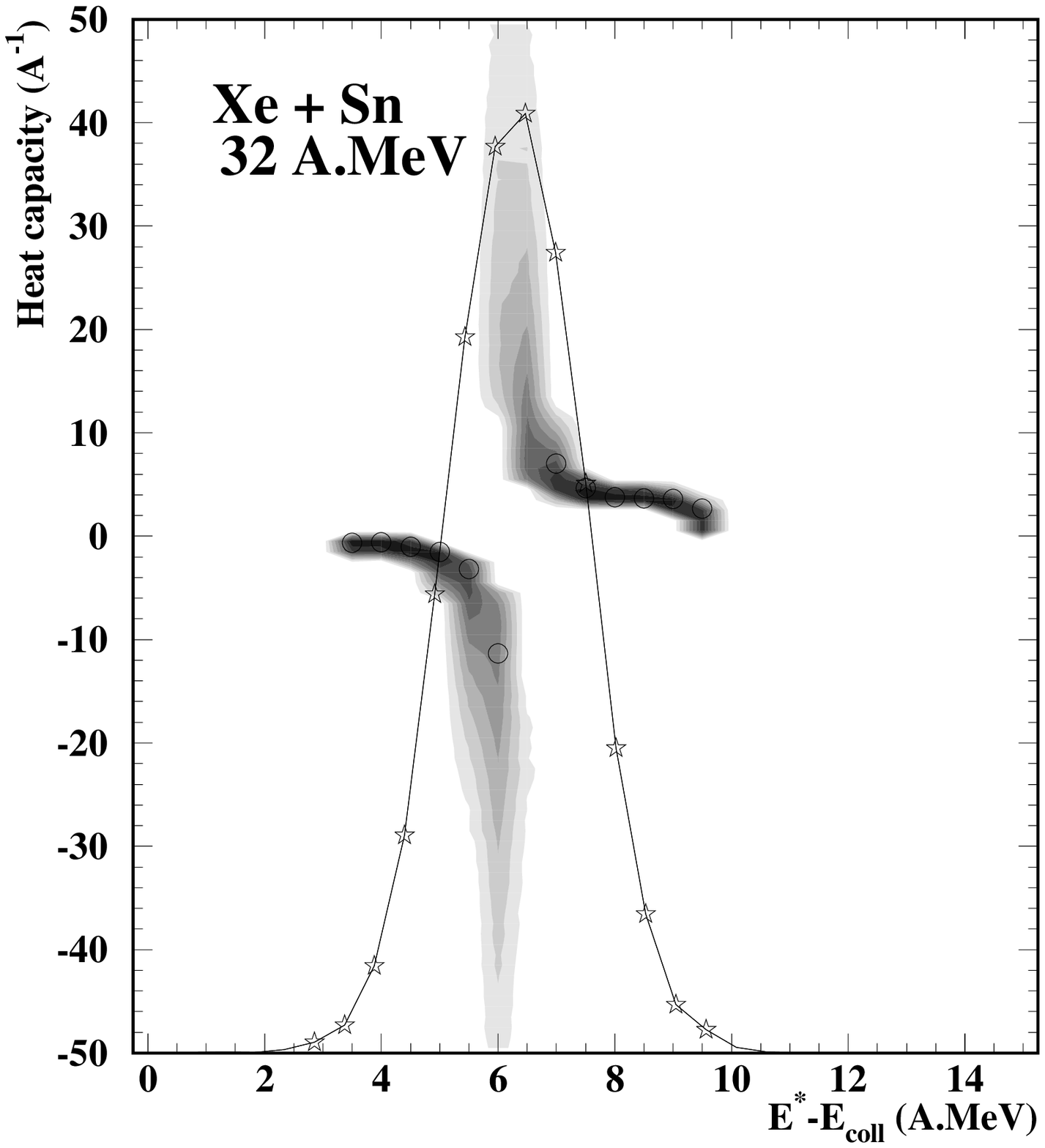}
\includegraphics[width=4cm]{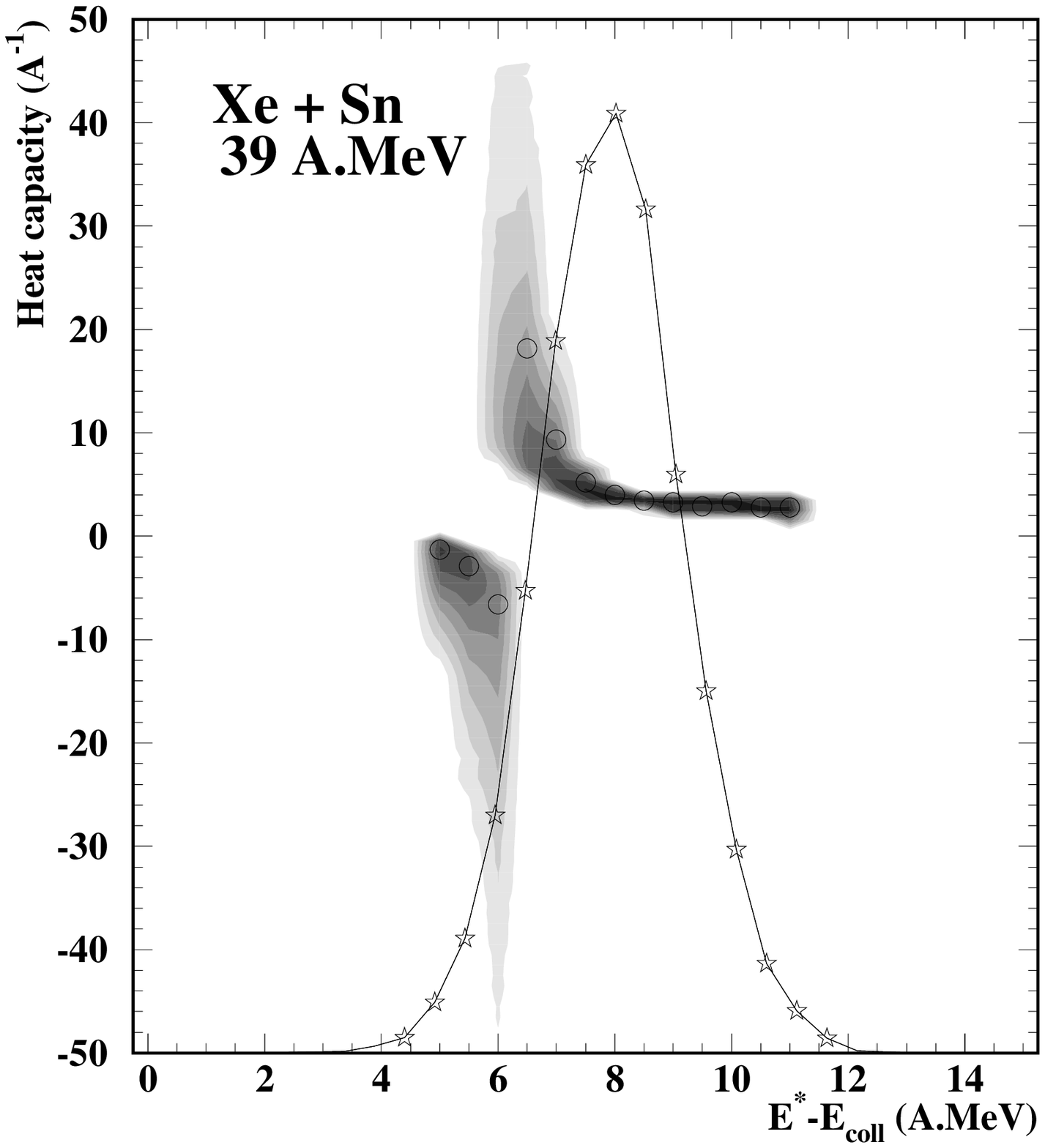} \\
 \includegraphics[width=4cm]{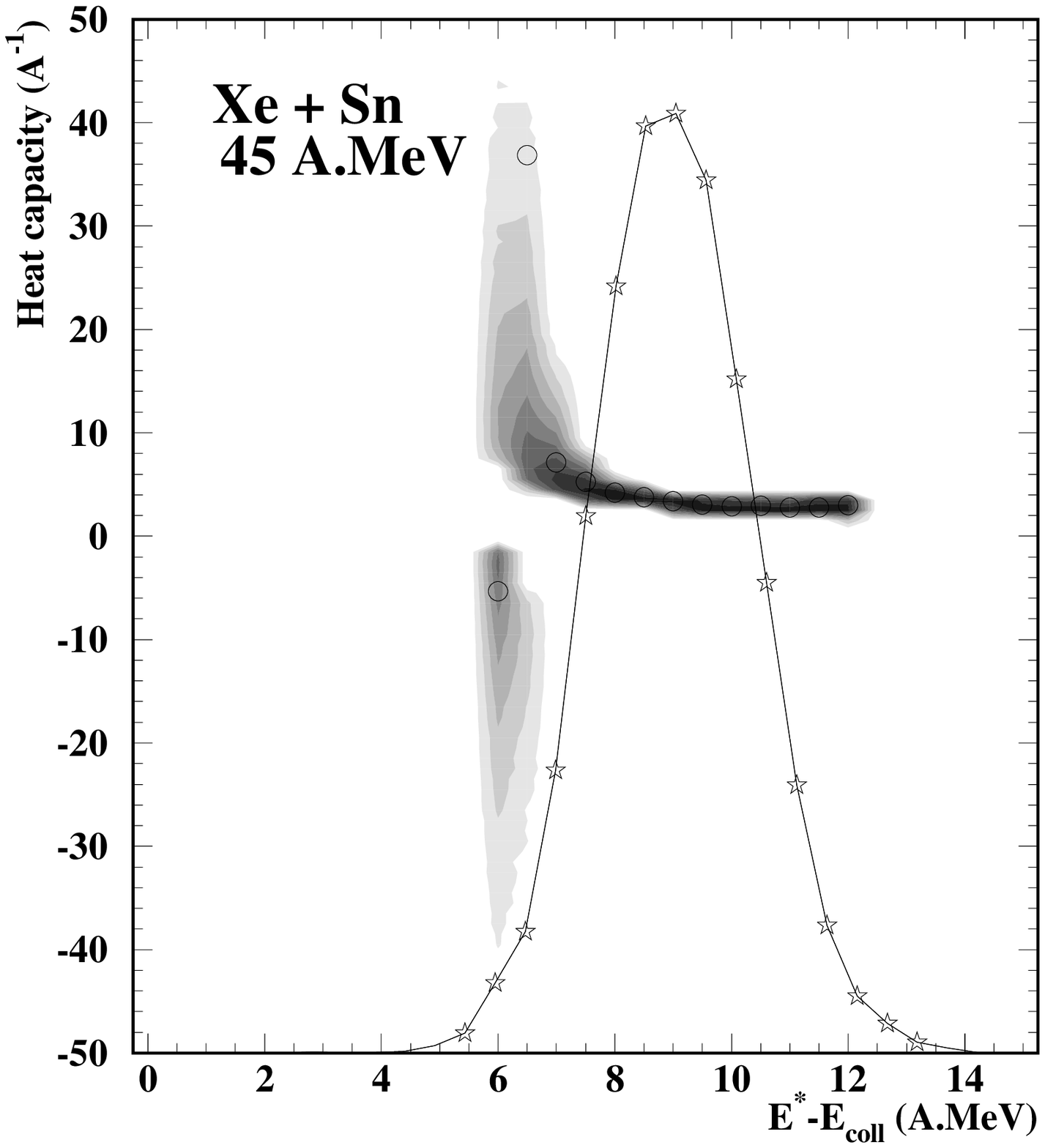}
\includegraphics[width=4cm]{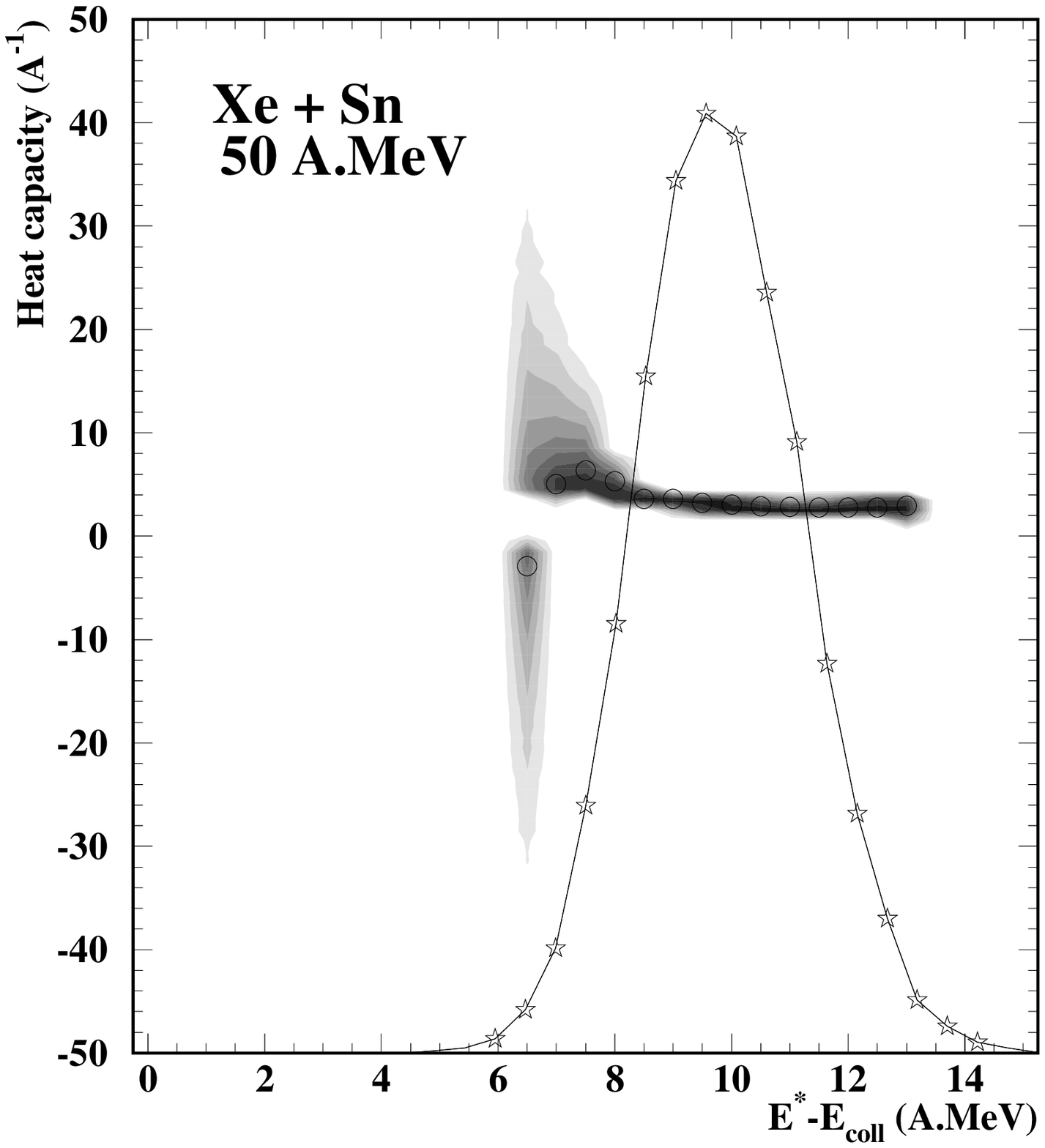}
\vspace*{-3mm} \caption{\footnotesize
Heat capacity versus the excitation energy (corrected for the
collective energy) measured in central collisions between Xe and Sn.
From~\cite{BouBorm00}} \label{cnegxesn}
\end{figure}
The  heat capacity was determined on reconstructed freeze-out configurations
out of the measured data~\cite{BouBorm00}. The excitation energy scale
was obtained by calorimetry, including all measured fragments and twice
the light charged particles emitted between 60 and 120$\deg$ in the c.m.
Neutrons were estimated from mass conservation. The
radial expansion energy was determined from a comparison of the data with
results of a statistical multifragmentation model, and was subtracted from
the excitation energy to give the energy distributions shown by the lines in
fig.~\ref{cnegxesn}. The potential energy was taken as the sum of the
mass balance and of the Coulomb energy estimated through the Wigner-Seitz
approximation, with a freeze-out volume $V=3 \times V_0$. The influence of
these assumptions on the amplitude of the fluctuations is thoroughly
discussed in ~\cite{MDA01}. The heat capacity, derived as explained in 
sect.~\ref{thermo}, is plotted as grey zones (error bars) in 
fig.~\ref{cnegxesn}. Negative branches are
observed below E$^*=$6-6.5~AMeV at the four incident energies. 
At 32 and 39 AMeV there is a significant overlap between the energy 
distribution and the negative heat capacity; this indicates that 
multifragmentation occurred in the spinodal region, in agreement with the
signal given by the charge correlations.
At and above 45 AMeV, both the non observation of enhanced equal-size
fragment partitions and the positive value of the heat capacity plead for
the system  multifragmenting in the coexistence region
outside the spinodal zone.

\vspace*{-5mm} \subsection{\textmd{\textit{The asymmetric entrance channel}}}

\vspace*{-3mm} 
\begin{figure}[!hbt]
\includegraphics*[height=7cm, width=8cm]{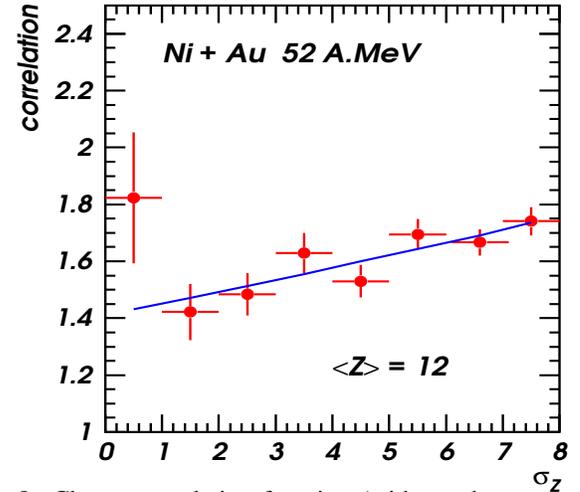}
\vspace*{-10mm} {\caption{\footnotesize Charge correlation function (without
charge conservation constraint) versus
the standard deviation $\sigma_Z$ for a given value Z$_{av}$ = 12. The line
indicates the linear background.From~\protect\cite{Guiot:ud}}} \label{corniau}
\end{figure}
Enhanced partitions of equal-sized fragments were looked for in central
Ni+Au collisions at 32 and 52 AMeV~\cite{guiot01}. A statistically 
significant  enhancement only appeared at 52 AMeV, as shown in
fig.~\ref{corniau}, where the correlation function is calculated without 
charge conservation constraint. Results with the new method can be found 
in the contribution of B.~Guiot, and agree with the above statement.

The heat capacity was calculated for incident energies of 32, 52, 74 
and 90 AMeV. Precisions on the implied assumptions are detailed in 
the contribution of O.~Lopez to this conference, in which the results 
are also unveiled.  The energy scale is here obtained from calorimetry, 
using products emitted between 60 and 120$\deg c.m.$  whatever their 
charge, due to stronger
contributions of forward preequilibrium emissions with respect to
the symmetric entrance channel. A negative branch of the heat capacity 
appears below 6-7 AMeV  at the four incident energies.
The data at 32 AMeV should be taken with caution, as the average 
completeness of the events is poor (65\% of the total charge). Indeed
because of the low value of the average excitation energy (3.5 AMeV), one 
would expect to observe the first divergence of the heat capacity, while 
only the second one is seen, in the higher part of the energy distribution. At
the three other incident energies, the average excitation energy varies
little (5.5, 7, 7 AMeV) and the heat capacity is negative in the first half
of the distribution.

For this asymmetric entrance channel, multifragmentation in the spinodal
region is shown by dynamical and thermodynamical
signals at 52 AMeV only. For collisions at higher energy only the
negative heat capacity was looked for, and found. At 32 AMeV there is no
sign of spinodal decomposition, and the data resemble more a 
fusion-evaporation process~\cite{BelBorm01}; surprisingly only the 
negative part of the heat capacity was observed despite the low excitation
energies covered. 

\section{\textit{New signals evidencing a phase transition}}

\vspace*{-3mm}
\subsection{\textmd{\textit{Universal fluctuations of an order parameter}}}

\vspace*{-3mm}
\begin{figure}[htp]
\includegraphics[scale=0.7]{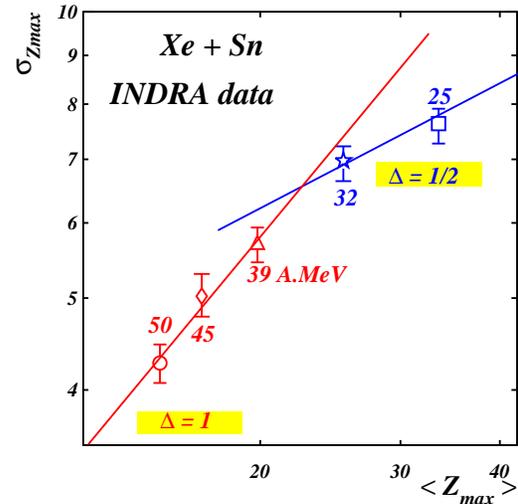}
\vspace*{-8mm} \caption{\footnotesize 
Normalised first and second moments of the scaled distributions of
the event heaviest fragments, for Xe+Sn central collisions at five incident
energies. The lines $\Delta$ = 1 and 1/2 are shown to guide the eye. Note
the two logarithmic scales. Adapted from~\protect\cite{Bot01}.} \label{flucuniv}
\end{figure}
The recently developed theory of universal fluctuations in finite systems
provides methods to characterize  critical and off-critical behaviours,
without any equilibrium assumption~\cite{Bot00}. In this framework, universal 
$\Delta$ scaling laws of one of the order parameters, p, should be observed:
 
$ \langle p \rangle^{\Delta} P(p) = 
\phi ((p - p^{mp})/ \langle p \rangle ^{\Delta}) $ \\
where $p^{mp}$ stands for the most probable value of p.
$\Delta$=1/2 corresponds to small fluctuations, 
$\sigma_p^2 \sim \langle p \rangle$, and
thus to an ordered phase. Conversely $\Delta$=1 occurs for the largest
fluctuations nature provides, $\sigma_p^2 \sim \langle p \rangle^2$, 
in a disordered phase.
 Scaling laws may be expected for an order  parameter p increasing with
time, for instance the fragment multiplicity in a fragmentation process,
or the largest fragment size in an aggregation process.
The method was applied to the Xe+Sn central collision samples described in
sect.~\ref{A250}, extended to the measurements performed at 25
AMeV~\cite{Bot01}. The
fragment multiplicity distributions at the five energies scale with a small
variance ($\Delta$=1/2), while the largest fragment distributions follow a
$\Delta$=1/2 scaling at low energy (25 and 32 AMeV) and a $\Delta$=1 scaling
for the three highest energies; fig.~\ref{flucuniv} illustrates the variation of
$\sigma_{Z_{max}}$ vs. $Z_{max}$. Following the authors of this study, this
indicates the transition from an ordered phase (evaporation?) to a
disordered phase, the fragments being produced following an aggregation
scenario. Note that spinodal decomposition well enters this type of scenario. 

\vspace*{-5mm} \subsection{\textmd{\textit{Bimodality of an order parameter}}}

\vspace*{-3mm} 
\begin{figure}[!hbt]
\includegraphics*[scale=0.7]{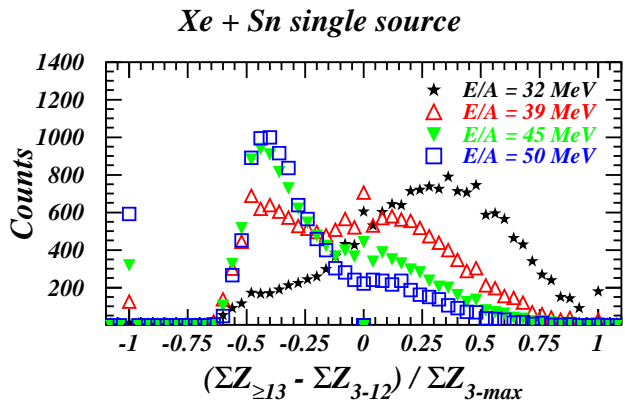}
\vspace*{-10mm} \caption{\footnotesize Distributions of the differences of 
total charges
bound in big (Z$\geq$13) and small ($Z \in [3-12]$) fragments in events from
Xe+Sn central collisions. From~\cite{BBorm01}} \label{bimod}
\end{figure}
An extended definition of first order phase transitions in finite systems
was proposed in~\cite{Cho01}, based on topology anomalies of the event
distribution in the space of observation. In this framework when a nuclear
system is in the coexistence region, the probability distribution of an order
parameter is bimodal, each component being a phase.
The Xe+Sn central collisions samples were also analysed this way. Having in
mind the liquid-gas coexistence region, one has to guess which fragments
belong to each of the phases. Owing to the detailed shape of the charge
distributions~\cite{BBorm01}, Z=12 was tentatively put as a limit between
the two phases. The chosen sorting parameter, 
$(\sum{Z_{Z \geq 13}} - \sum{Z_{Z \in [3-12]}}) / \sum{Z_{Z \geq 3}}$, 
may be connected with the density difference of the two phases 
$(\rho_L - \rho_G)$, which is an order parameter~\cite{Gul01}. The results
are displayed in fig.~\ref{bimod}: none of the distribution is Gaussian,
which would sign a pure phase; but the gas phase is dominant above 45 AMeV,
while the liquid phase is more important at 32 AMeV. The two
components are roughly equal at 39 AMeV. 

Similar results were obtained on the Ni+Au central collision samples selected
with the PCA. A similar sorting parameter, making use of only the three largest
fragments, was used. Its probability distribution exhibits two nearly equal 
peaks at 32 AMeV, a big ``gas'' peak with a small shoulder at 52 AMeV 
and a single phase picture  at 90 AMeV~\cite{LopBorm01}.

\section{SYNTHESIS OF THE OBSERVATIONS FOR SYSTEMS OF 250 NUCLEONS}

\vspace*{-3mm} 
\begingroup \squeezetable
\begin{table*}[!hbt]
\caption{\footnotesize{Summary of the phase transition signals described in
the previous sections.
$\varepsilon_{inc}$, $\varepsilon_{th}$, $\varepsilon_{rad}$, stand
for the incident, thermal and expansion energies in AMeV (see text for
precisions). SD means signal of spinodal
decomposition. $\Delta$ indicates the exponent of the universal scaling, and
the Bimod line gives the number of peaks in the probability
distribution (\textcircled{2} means two equal peaks).
A ``-'' means that no measurement of the concerned observable was done.}}
\begin{ruledtabular}
\begin{tabular}{c|cccccccc}
system & Ni+Au & Xe+Sn & Xe+Sn & Ni+Au & Xe+Sn & Xe+Sn & Ni+Au &  Ni+Au \\
$\varepsilon _{inc}$ & 32. & 32. & 39. & 52. & 45. & 50. & 74. & 90. \\
\hline
$\varepsilon _{th}$ & 5.0 & 5.0 & 6.0 & 6.5 & 6.6 & 7.0 & 7.0 & 7.5 \\
$\varepsilon _{rad}$ & 0. & 0.6 & 1.2 & 0. & 1.7 & 2.0 & 0. & $\leq$ 0.5 \\
\hline
SD & no & yes & yes & yes & no & no & - & - \\
c$<$0 & yes &  yes  & yes  & yes & no & no & yes  & yes \\
$\Delta$ & - & 1/2 & 1 & - & 1 & 1 & - & - \\
Bimod. & \textcircled{2} & 2 & \textcircled{2} & 1 or 2 & 2 & 2 & - & 1 \\
\end{tabular}
\end{ruledtabular}
\end{table*} \endgroup

A synthesis of the signals which may sign a liquid-gas phase transition for
nuclei of mass $\sim$250 is presented in table I.
The thermal and radial energy scales come from SMM simulations, with
respective uncertainties of 1 and 0.5 AMeV, depending, for instance,
on the event selections, the freeze-out volume chosen {\ldots}
~\cite{BouBorm00,BelBorm01}. The first evidence coming from this table is
that for the asymmetric entrance channel, multifragmentation of the fused
system formed in central collision is \emph{not} associated to a radial
expansion. This comes at variance of other results for similar
systems in the same energy range, for instance Kr+Au~\cite{Wil97}. Note that
the result comes directly out of the data, as for Xe+Sn at 50 AMeV and Ni+Au
at 90 AMeV, the partitions are similar, while the average kinetic energies
of a given fragment strongly differ, and are constant for Z between 12 and
30 for the latter system~\cite{LopBorm01}

For both entrance channels, the systems reach the spinodal region and remain 
there for a time sufficient to allow the development of spinodal 
instabilities for
thermal energies between 5 and 6.5 AMeV, as attested by the observation of
favoured equal-sized fragment partitions and the negative values of the heat
capacities. When the thermal energy increases, the expansion energy seems to
have a decisive role: without expansion, the picture is similar
to the one just described, up to $\varepsilon _{th}=$7.5 AMeV,
while an expansion energy above 25\% of the thermal energy would
make the system cross too fast the spinodal region, toward the
coexistence region or even the gas region. In this case neither the spinodal
decomposition nor the negative heat capacity are observed. 
The critical excitation energy, given by the maximum thermal energy of 
the systems described in this paper and
shown to have fragmented in the spinodal region is at least 7.5 AMeV, much
above the value given in \cite{Ell01}. 
It must however be noted that the ``multifragmentation'' partitions analysed 
in this reference comprise essentially one large 
cluster and one intermediate mass fragment, together with nucleons and light 
charged particles~\cite{Bea01}, at variance with the partitions observed in
the INDRA data which have a larger number of fragments. In this sense 
the ``critical'' values of \cite{Ell01} would rather be assimilated to 
``limiting'' temperatures, or excitation energies, as described 
in~\cite{Lev85}. Above limiting values, a
nucleus is unstable due to Coulomb forces, and can only de-excite through
multifragmentation.

What additional information is brought by the other signals quoted in 
table I? The universal
fluctuations are small for $\varepsilon _{th}\leq$~5 AMeV, and large above
($\Delta=$~1). This can be interpreted as the transition from the dominance of
fusion evaporation to that of fusion-multifragmentation processes around
this energy; indeed at 32 AMeV the selected central collisions in Ni+Au
present essentially the characteristics of an evaporation
process~\cite{BelBorm01}. In this sense, large fluctuations of Z$_{max}$
would sign the break-up of the system inside the coexistence region. 
Bimodality of a parameter related to the density of the system goes roughly
in the same direction. Two peaks, indicating that the systems break in the
coexistence region, are observed in
all the cases except for Ni+Au at 90 AMeV. This would indicate a pure gas
phase, and contradict somewhat the measurement of a negative heat capacity. 
The equal abundance of the two phases occurs for 
$\varepsilon _{th}\leq$~5-6 AMeV, above that energy nuclear systems would
mainly multifragment. 

\section{CONCLUSIONS AND PERSPECTIVES}

\vspace*{-3mm} The results obtained in analysing central collisions leading 
to A$\sim$250
systems through a symmetric and an asymmetric entrance channel show several
features which may be taken as characteristics of a liquid-gas phase
transition in nuclei. While each of these signals may not be conclusive by
itself, their concomitance gives strength to the assumed scenario. In the
next future, one may try to find other variables or signals which would
support the existence of the phase transition. It is also worth looking to
all these signals for quasi-projectile nuclei, of which INDRA possess a
tremendous collection, spanning Ar, Ni, Xe, Gd, Ta, U projectiles. 

 Next, enriched knowledge of nuclear matter properties will come from taking
into account its two component nature (neutrons and protons), by exploring
collisions induced by exotic projectiles when they are available. Isospin
fractionation should allow to distinguish the liquid and gas phases.

\begin{acknowledgments}
\vspace*{-3mm}
All the exciting results presented in this paper were understood through 
their sharing  with theoreticians. 
The INDRA collaboration is indebted to P.~Chomaz, M.~Colonna, P.~Désesquelles,
F.~Gulminelli and M.P{\l}oszajczak for their essential contribution through 
many discussions.
\end{acknowledgments}

%\bibliography{mftp}
%% Faire un premier passage latex. ensuite taper ``bibtex iwm2001'', qui
% utilise le fichier iwm2001.aux cree. puis refaire un latex - ou deux

\end{document}